# Limits to inertial vibration power harvesting: power-spectral-density approach and its applications

**Akshay Ananthakrishnan[1], Inna Kozinsky[2] and Igor Bargatin[1]**
[1] Department of Mechanical Engineering and Applied Mechanics, University of Pennsylvania, Philadelphia, PA, USA, 19104
[2] Research and Technology Center, Robert Bosch LLC, Palo Alto, CA 94304

E-mail: bargatin@seas.upenn.edu

**Abstract:** Maximum output powers of vibration-driven inertial power harvesters reported in literature exhibit sizable variations, even when normalized by the device weight or their maximum linear size. To help establish a common benchmark, we present a power-spectral-density based approach for estimating the maximum power that can be obtained using a resonant inertial power harvester from a random (aperiodic) vibration source with a given power spectral density. In the simplest case of unlimited harvester size, the maximum obtainable power is simply proportional to the maximum value of the power spectral density of vibration acceleration. We describe in detail the underlying theory and the practical method for evaluating these limits. We also present a simple analytical formula to estimate the minimum harvester size required for obtaining the maximum possible power. Specific power limits are derived as function of harvester size for three practical examples of vibration sources: (a) pneumatic power tool, (b) the body of an idling Mazda RX7 sports car, and (c) human walking motion. Characteristic power spectra and optimum design parameters (quality factor and resonant frequency) are presented for both translational and rotational harvesters. Translational harvesters generally outperform rotational ones for realistic harvester sizes, with the power tool vibrations yielding a practical power limit of ~300 mW per gram of inertial mass, followed by walking at ~1mW/g, while the vibrations of a car body yield ~0.1 mW/g or less.

## 1. Introduction

The need to power a myriad of portable devices continues to drive the pursuit for alternative energy sources. While batteries remain the preferred power source for most devices, frequent charging and replacement requirements render them expensive or impractical in some cases, such as key fobs, pacemakers, *etc*.[1,2]. Power harvesters capable of scavenging energy from the ambient could offer a promising alternative.

Vibration power harvesters are typically designed to be powered by ambient vibrations, which are ubiquitous and allow a variety of applications ranging from industrial machinery[1,3,4,5] to integrated wireless sensor nodes and micro systems[3,6,7,8,9,10]. Many power harvesters have been developed specifically to utilize the movements of a human body [11] using electromagnetic [12,13], , electrostatic

[14] or piezoelectric [5,15,16,17] power conversion mechanisms. Power outputs from such devices vary depending on the actuation mechanism and location on the human body and typically range from a few hundred microwatt for inertially driven vibration power harvesters [8,18] to several hundred milliwatt for impact-based harvesters [12,15]. Several bio-compatible versions of such harvesters have been developed to power implantable biomedical devices and health monitoring systems [2,11,19,20,21]. Vibration power harvesters have also been considered for automotive applications as an integral part of hydraulic engine mounts [22] and regenerative suspension systems [4,23], with the latter estimated to yield up to 2000 Watt during driving [24].

To maximize output power, resonant powers harvesters are typically tuned to be in resonance with one of the prevailing frequencies of the ambient vibrations [25]. For resonant devices excited by stochastic translational vibrations, the output power depends crucially on harvester size, its operating frequency and bandwidth at resonance [26,27], which must be chosen carefully to optimize the power output. In addition, techniques like frequency up-conversion [28,29,30], non-linear device design [31,32,33], dynamic resonance tuning [25,27] and bandwidth widening [4,34] can be employed to optimize the harvesting of vibrations and produce maximum power. However, practical limitations on harvester sizes place an upper bound on the maximum displacement of proof mass, thereby limiting the maximum obtainable power [1,33]. These limitations can potentially be alleviated by employing a rotating proof mass architecture [35], which relies on rotational vibrations and therefore has a different dependence on device size.

Since heavier inertial power harvesters are expected to produce more power, it is convenient to normalize the output power by the mass of the harvester. Such normalized power outputs still exhibit large variations across demonstrated vibrational power harvesters, depending on their designs, transduction principles, and performance characteristics. For instance, an electromagnetic inertial harvester developed by Zhang et al. [13] generated a maximum power of ~0.2 mW/gram of harvester mass, when placed in a backpack under slow running conditions. Another inertial electromagnetic harvester developed by Saha [12] yielded maximum power output of ~0.1 mW/g of harvester mass, when positioned in a rucksack under similar running conditions. For comparison, the directly actuated (i.e., not inertial) knee-mounted generator developed by Neill Elvin and Alex Elvin [36] produced power output of ~1 mW/g of harvester mass under optimally tuned conditions at a reasonable walking pace.

There is a clear need for a common benchmark for the power produced by an inertial power harvester form a given vibration source. Having such an established benchmark would allow comparisons of the relative efficiency of different designs of inertial harvesters driven by random vibrations. Recently, Buren et al. [37] used time-domain simulations of the motion of the proof mass to calculate the maximum theoretical output power of inertial generators for the particular case of walking or running vibrations. They obtained power limits of up to ~1 mW depending on the location on the body and the size of the harvester (Fig. 7 of Ref. [37])

This paper presents a method to estimate the maximum power output of an inertial power harvesters that is driven by *any* random vibration source with a given power spectral density (PSD). Section 2 introduces the theoretical aspects of this approach and describes the underlying analytical model. Section 3 presents limits for vibration power harvesting from three specific sources (a) power tools (b) cars (c) human walking motion, which were derived using the proposed method. The scope of this paper is limited to linear velocity-damped vibration power harvesters, whose spectral response can be described by the standard complex Lorenzian. Nonlinear power harvesters which can theoretically exceed these limits, but require more complex designs and analysis, are not discussed in this paper. We also assume that the operating harvester does not affect the vibration source, e.g., a human-motion power harvester should not significantly change the way you walk when wearing a harvester.

2. **Theory**

Ambient vibrations are typically random and cannot be described adequately by a sine wave, *i.e.*, a delta-function power spectral density (PSD). The behavior and response of linear vibration harvesters to broadband excitations with a given PSD has been considered previously by Halvorsen [38]. However, the corresponding power limits were derived in a rather general and abstract way and their significance for practical applications was not entirely obvious. In this section, we introduce the relevant notation and re-derive PSD-based expressions in a less general but simpler way.

We first analyze a linear, translational type vibration harvester operating at resonant conditions. The steady-state response, *i.e.*, vibration amplitude, $x$, of such a device excited by vibration-induced, periodic acceleration, $a$, at a frequency, $\omega$, is given by the standard complex Lorentzian function:

$$x(\omega) = a / (\omega_0^2 - \omega^2 + i\gamma\omega) \tag{1}$$

Where $\omega_0$ is the resonant frequency in rad/sec, and $\gamma = \omega_0/Q$ is the resonance line width or dissipation constant corresponding to the resonator's quality factor, $Q$.

Before introducing complexities associated with stochastic nature of vibrations, we consider a simple case where the inertial resonator is driven by white noise. This enables us to deal with a simple one-sided PSD with a constant amplitude, $\hat{S}_a(\omega) = \hat{S}_a$, where frequency, $\omega$, is denoted in units of rad/sec. The displacement of an inertially driven resonator for this specific case is characterized by the following PSD:

$$\hat{S}_x(\omega) = \hat{S}_a / [(\omega_0^2 - \omega^2)^2 + \gamma^2 \omega^2] \tag{2}$$

For harmonic excitation, the instantaneous power, $P$, dissipated by the resonator of mass, $m$, with an instantaneous velocity, $v$, and quality factor, $Q$, is given by:

$$P = m\gamma v^2 = m\omega_0 v^2 / Q \tag{3}$$

However, in practice, not all of the dissipated power, $P$, can be utilized to generate electricity, since a portion of it is always lost to the ambient as heat. Correspondingly, the total dissipation constant of the harvester, $\gamma$, can be written as $\gamma = \gamma_h + \gamma_e$, where the heat-producing dissipation, $\gamma_h$, describes the irreversible loss of energy to heat through various forms of mechanical friction, resistive losses, etc., and the electrically induced dissipation, $\gamma_e$, describes the conversion of the mechanical energy into useful electrical power. Assuming that the heat-producing loss is negligible, i.e., $\gamma \approx \gamma_e$, and all of the dissipated power is efficiently collected as useful electrical power, we can use fundamental principles of Fourier transforms to obtain the maximum electric power output:

$$P = m\gamma\langle v^2 \rangle = m \int_0^\infty \gamma \omega^2 \hat{S}_x(\omega) d\omega \tag{4}$$

Substituting $\hat{S}_x(\omega)$ as obtained from Eq.2 and performing a change of variable using $z = \omega/\gamma$ we obtain:

$$P = m\hat{S}_a \int_0^\infty z^2 / [(Q^2 - z^2)^2 + z^2] \, dz \tag{5}$$

Solving the above integral using standard methods for rational fractions, we obtain the following result

$$P = \pi m \hat{S}_a/2 = mS_a/4 \tag{6}$$

Where $\hat{S}_a$ denotes PSD using angular frequency ($\omega$) units of rad/sec, and $S_a$ denotes PSD using cyclic frequency ($f$) units of Hz. These two terms are coupled through the simple relation $\hat{S}_a(\omega) = S_a(f)/2\pi$. Notably, the maximum power does not depend on the resonator's $Q$ value. While higher quality factors lead to higher oscillation amplitudes on resonance, they also narrow the range of frequencies for which the device remains in resonance, *i.e.*, lead to narrower bandwidth. The two effects cancel each other, resulting in power output that is independent of the quality factor. We note that equation (6) differs by a factor of 2 from equation (19) in Ref.[38] because we use a one-sided PSD (frequencies are always positive) instead of the two-sided PSD (both negative and positive frequencies allowed) in Ref. [38]. Practical vibration power harvesters are not usually driven by white noise, but rather by vibrations with and arbitrary PSD, schematically represented in figure 1.

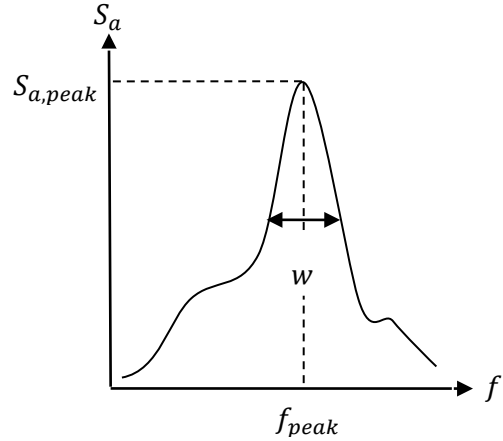

**Figure 1.** Schematic of an arbitrary PSD.

Given the peak PSD value, $S_{a,peak}$, which occurs at frequency $f_{peak}$, we have $S_a(f) \leq S_{a,peak}$ or, equivalently, $\hat{S}_a(\omega) \leq \hat{S}_{a,peak} = S_{a,peak}/(2\pi)$ and the maximum power that can obtained from the harvester is given by the inequality:

$$P = m\int_0^\infty \hat{S}_a(\omega)\gamma\omega^2/[(\omega_0^2 - \omega^2)^2 + \gamma^2\omega^2]d\omega$$

$$\leq m\hat{S}_{a,peak}\int_0^\infty \gamma\omega^2/[(\omega_0^2 - \omega^2)^2 + \gamma^2\omega^2]d\omega \tag{7}$$

Based on prior analysis of the white noise PSD, we notice that RHS term in the above inequality reduces to the term $\pi m\hat{S}_{a,peak}/2$ in accordance with equation (6). The output power given by equation (7) can then be rewritten as:

$$P = m\int_0^\infty \hat{S}_a(\omega)\gamma\omega^2/[(\omega_0^2 - \omega^2)^2 + \gamma^2\omega^2]d\omega$$

$$\leq \quad \pi m \hat{S}_{a,peak}/2 = m S_{a,peak}/4 \tag{8}$$

If the PSD is a smooth function near the peak, $[\omega_{peak}, \hat{S}_{a,peak}]$, this upper bound can be asymptotically reached by using resonators with the resonance frequency $\omega_0 = \omega_{peak}$ and increasingly high $Q$. The spectral response of such high quality factor resonators is very narrow and can be approximated by a delta function, *i.e.* $\gamma \omega^2 / ((\omega_0^2 - \omega^2)^2 + \gamma^2 \omega^2) \approx \pi \delta(\omega - \omega_0)/2$ as $\gamma \to 0$. The maximum power output from a high-$Q$ harvester then approaches the limits given by the RHS of equation (8):

$$\lim_{\substack{\omega \to \omega_0 \\ Q \to \infty}} P = \pi m \hat{S}_{a,peak}/2 = m S_{a,peak}/4 \tag{9}$$

In practice however, the acceleration PSD exhibits a characteristic width, $w$, usually measured as the full width at half maximum (FWHM) illustrated in figure 1. In order to yet approach the maximum power level asymptotically, it is sufficient to use a resonator with $Q \gg f_{peak}/w$.

We note that using a high $Q$ factor leads to large displacement of inertial mass at resonance, which in turn necessitates a larger resonant harvester. Practical applications require reasonably sized harvesters and thus may restrict maximum displacements and $Q$ values. In order to determine influence of size on allowed maximum displacement values, dispersion of equation (2) can be calculated as:

$$\langle x^2 \rangle = \int_0^\infty \hat{S}_x(\omega)\, d\omega$$

$$= \int_0^\infty \hat{S}_a(\omega)/[(\omega_0^2 - \omega^2)^2 + \gamma^2 \omega^2]\, d\omega \tag{10}$$

For resonators with $Q \gg f_{peak}/w$, we can simplify the above integral to:

$$\langle x^2 \rangle \cong \pi \hat{S}_{a,peak}/2\gamma \omega_{peak}^2 = \pi Q \hat{S}_{a,peak}/2\omega_{peak}^3 \tag{11}$$

The standard deviation of displacement, $\sigma_x$, can then expressed as:

$$\sigma_x = \sqrt{\langle x^2 \rangle} = \sqrt{\pi Q \hat{S}_{a,peak}/2\omega_{peak}^3} \tag{12}$$

In practical situations, randomness of the excitation may lead to maximum displacement overshooting the standard deviation value calculated in equation (12). Choosing a safety factor of $n$, the probability that $|x| > n\sigma_x$ is given by $1 - \text{Erf}(n/\sqrt{2})$, for Gaussian noise and is less than $1/n^2$ for any probability distribution according to Chebyshev inequality. Accordingly, for $n = 3$ the probability of $|x| > 3\sigma_x$ is less than 1% for Gaussian noise and 11.1% in the general case. .Since the inertial mass can move to either side of the equilibrium position, the minimum harvester size is given by ($n = 3$):

$$L \geq 2n\sigma_x = 30\sqrt{\pi Q \hat{S}_{a,peak}/2\omega_{peak}^3} = \frac{15}{2\pi f_{peak}}\sqrt{Q S_{a,peak}/(2\pi f_{peak})} \tag{13}$$

Equation (13) determines the allowed combinations of $Q$ and $\omega_{max}$ for a given harvester size, $L$. The

output power will saturate when $Q \gg f_{peak}/w$ and therefore when the harvester size satisfies

$$L \gg \frac{3}{2\pi f_{peak}}\sqrt{S_{a,peak}/(2\pi w)} \approx 0.19 \frac{\sqrt{S_{a,peak}/w}}{f_{peak}} \tag{14}$$

As we show below, in practice the output power saturates for sizes that are several times larger than the R.H.S. of equation (14), and it therefore convenient to introduce a critical harvester size $L_c$ using the following simple formula:

$$L_c = \frac{\sqrt{S_{a,peak}/w}}{f_{peak}} \tag{15}$$

We note that the dependence of maximum allowed $Q$ on harvester size can be eliminated by employing a rotational design for vibration power harvesters. Unlike their translational counterparts, these harvesters use of a rotating proof mass, *e.g.*, a ring, which is driven by rotational acceleration, $\alpha$, and which can in principle undergo infinite angular rotations. It is therefore possible to use such harvesters with arbitrarily high $Q$ values and operate them at the maximum power point $(f_{peak}, S_{\alpha,peak})$ shown in figure 1. The maximum power from an ideal rotational vibration power harvester, $P_r$, with moment of inertia, $I$, can then be derived similarly to equation (9) and is given by:

$$P_r = \pi I \hat{S}_{\alpha,peak}/2 = I S_{\alpha,peak}/4 \tag{16}$$

where $S_\alpha(f)$ is the PSD of rotational acceleration caused by the ambient vibrations.

## 3. Results

In this section, we use the PSD–based analysis to deriving power limits for inertial power harvesters driven by three specific vibration sources: a power tool, an idling car, and a walking person. In each of these three cases, the vibration-induced acceleration signal was recorded using a suitable three axis accelerometer. A fast Fourier transform (FFT) was then performed on recorded signals, and a one-sided PSD estimate was then obtained for each component using the standard Welch method. The signal component that has the highest peak value in PSD offers the maximum potential for power harvesting and was therefore isolated for further analysis.

To judge the applicability and usefulness of the limits derived above, we also calculated the output power and displacement numerically for a wide range of resonator frequencies and quality factors. The frequency of the resonator $f_0$, was varied linearly in the same range as the corresponding PSD data. The quality factor, $Q$, of the resonator was varied logarithmically in the realistic range from 0.01 to 1000. The output power, $P$, and corresponding average proof mass displacements, $\sigma_x$, were calculated using equation (8) and equation (10), respectively. The maximum proof mass displacement, $\sigma_x$, was limited to range from 0.1 µm to 10 m. Possible harvester sizes, $L$, thus ranged from 0.6 µm to 60 m in accordance with equation (13). While 60 m is an unrealistically large harvester size for most applications, we included such large sizes in the calculations to illustrate how the output power saturates at large harvester size.

To calculate the maximum power that can be produced by a harvester of given size, $L$, we swept the values of $Q$ and $f_0 = 2\pi\omega_0$ and permitted only those combinations that satisfied the condition on displacement $\sigma_x < L/6$ (see equation (13)) . The power values calculated for this restricted set of $Q$ and $f_0$ using equation (8) was then searched for the largest output power $P_{max}(L)$. This value sets the conditional power limit, $P_c(L)$, for a harvester that is subject to the condition $\sigma_x < L/6$. The corresponding $Q_{opt}(L)$ and $f_{opt}(L)$ are the optimal quality factor and resonant frequency. This process was repeated to determine power limits $P_{max}(L)$ for the three vibration sources considered.

*3.1. Power tool vibrations*
Pneumatic power tools such as hammers, drills, grinders, *etc.*, often produce high-frequency vibrations during their operation. For the purpose of analysis, vibrations of a small pneumatic hammer were recorded using a high-frequency three-axis accelerometer. With a sampling frequency of about 40.96 kHz, a time record of the signal consisting of 1300000 data points was generated. Figure 2 shows the Welch-averaged PSD, which features two peaks of almost equal height at frequencies of approximately 572 Hz and 498 Hz.

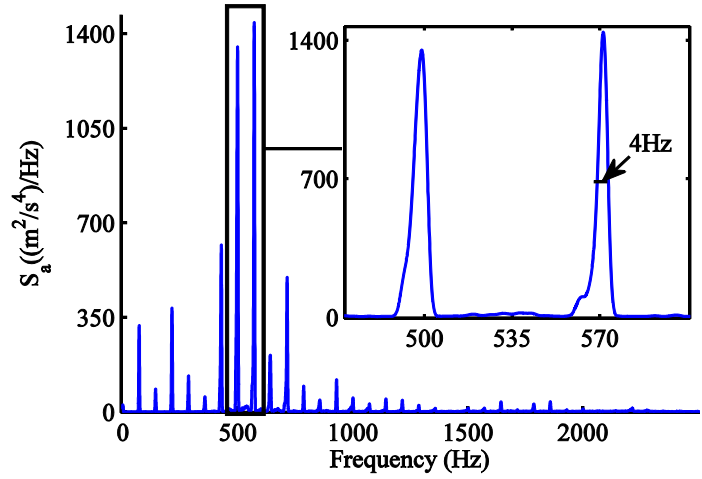

**Figure 2.** PSD of a small power tool.

Figure 3 is a log-log plot depicting the variations of $Q_{opt}(L)$ and $f_{opt}(L)$ with harvester size, $L$, while Figure 4 shows an expected overall increasing trend in conditional power, $P_c(L)$. The optimum frequency $f_{opt}(L)$ oscillates between 498 Hz and 572 Hz, but eventually settles at the frequency of the highest peak (572 Hz). Larger harvesters allow higher peak displacements and therefore higher quality factors, $Q_{opt}$, which is also reflected in Figure 3. For sufficiently large harvesters ($L > 1$ cm) operating at highest peak frequency $f_{opt} = 572$ Hz and the maximum allowed $Q_{opt} = 1000$, the maximum power saturates at $P_{max} = 313\ mW/g$. This agrees well with the limit given by equation (9) of $P/m = S_{a,peak}/4 \approx 1400/4\ m^2/s^3 = 350\ m^2/s^3 = 350\ W/kg = 350\ mW/g$. The small discrepancy of approximately 10% is explained by the fact that the PSD used for analysis is necessarily an estimate and varies depending on the particular parameters used for Welch averaging. As seen from Figure 4, the size at which the output power saturates, agrees within a factor of 2 with the estimate of equation (15):

$$L_c = (1/f_{peak})\sqrt{S_{a,peak}/w} = (1/572)\sqrt{1400/4} \approx 33\ mm$$

From figure 4 we also see that power output of translational harvester of size $L_c = 33mm$ is exactly equal to saturation power of 313mW/g. This gives us confidence that $L_c$, as calculated from equation (15), provides a good estimate of harvester sizes required to obtain maximum power.

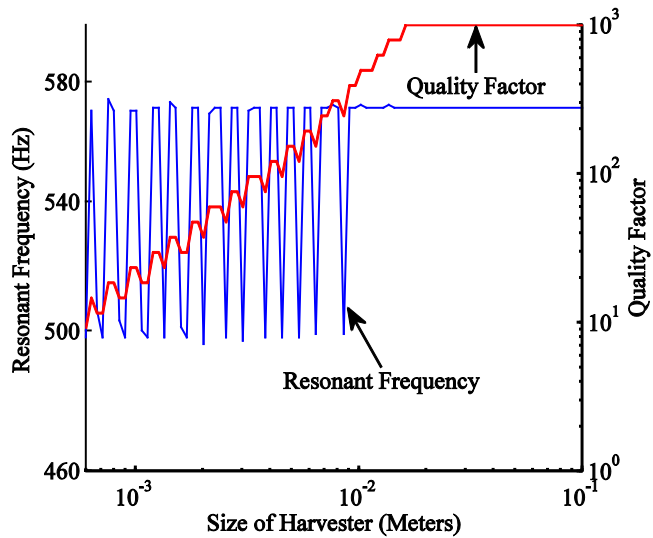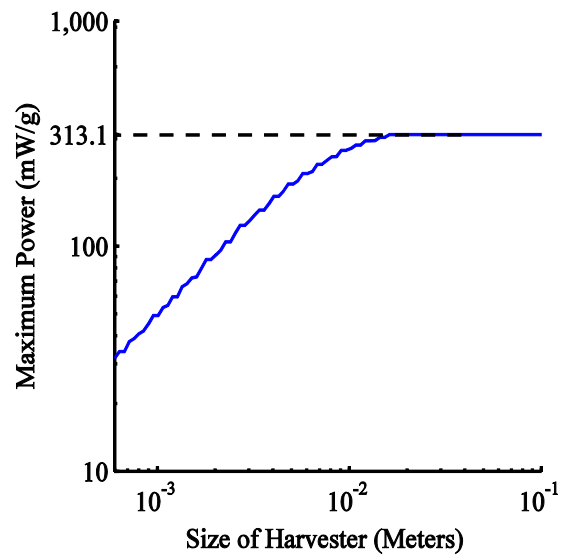

**Figure 3.** Optimum frequency and quality factor for a power-tool based harvester

**Figure 4.** Conditional power from power tool based harvester.

*3.2. Car vibrations*
Significant vibrations are present inside a car due to the noise produced by working engines, compressors, *etc*. as well as irregularities in terrain when driving. Figure 5 shows the Welch averaged PSD for vibrations recorded at two locations of an idling Mazda RX7 sports car: (1) passenger side firewall in engine bay and (2) base of shock tower/wheel hub of front passenger side wheel. For comparison, figure 5 also indicates an industry standard for vibration limits in vehicle locations with and without suspension [39]. The measured vibrations are lower than the corresponding standards by about two orders of magnitude, which is explained by the facts that the standards are conservative and that car was idling in a stationary position rather than driven at a high speed on a realistic terrain during the measurement.

Three-axis accelerometers were used to obtain one hundred time-records for each location, each lasting nearly 2.5 sec at a sampling frequency of about 2048Hz. For each location, the acceleration component yielding highest peak PSD value was selected and its time records were concatenated after eliminating mean values, i.e., the drift-induced DC offset in frequency spectrum.

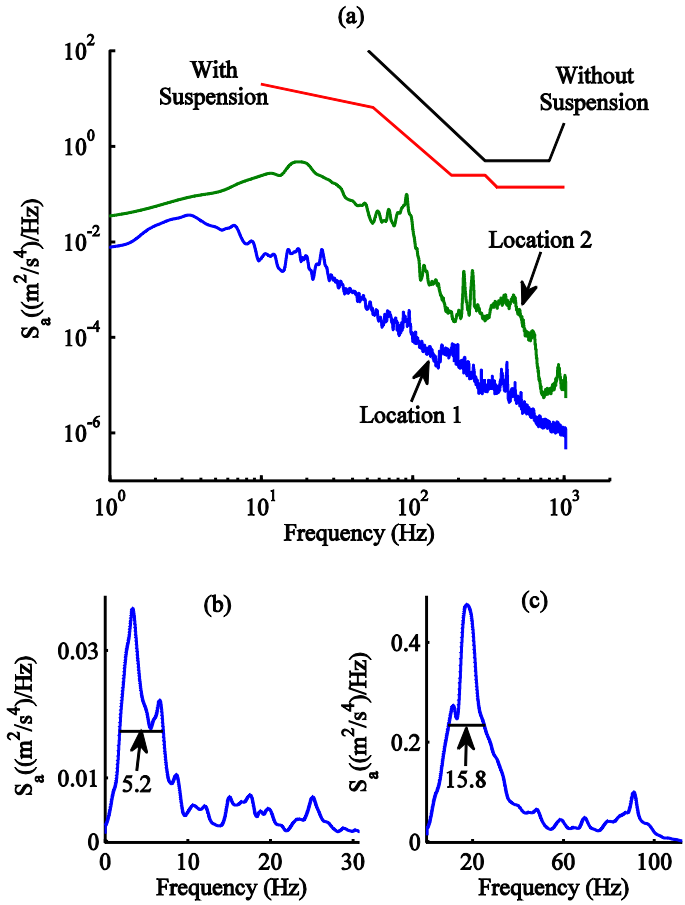

**Figure 3.** (a) PSD measurements of car vibrations compared to the industry standards [39] for automotive vibrations with and without suspension (b) PSD at location 1 zoomed in on highest peak (c) PSD at location 2 zoomed in on highest peak.

The total concatenated time record has a total of about 5 million data points and the time duration of $T = 250s$, yielding a PSD frequency step $\Delta f = 1/T = 4\ mHz$. The same frequency step size is therefore used in estimating integral expressions. In order to approximate the integrals in equation (8) and equation (10) as sums with a finite step size $\Delta f$, there must be a reasonable number of spectral points that lie within the width of the resonator peak, $f_{peak}/Q$. Hence, the following condition should be satisfied by frequency step size, $\Delta f$:

$$\Delta f \ll w = f_{peak}/Q \text{ or } Q \ll f_{peak}/\Delta f \tag{17}$$

For the broad peak centered around $f_{peak} = 3\ Hz$ in Fig.5, we obtain $Q \ll f_{peak}/\Delta f = 3\ Hz/4\ mHz = 750$. Therefore, in this calculation, the quality factors ranged from 0.01 to 100 instead of the previously used range from 0.01 to 1000 to ensure reasonable accuracy of integral expressions in equation (8) and equation (10).

Plots demonstrating $Q_{opt}(L)$ and $f_{opt}(L)$, for both locations 1 & 2 are shown in figure 6 and figure 7 respectively. As expected, the highest $Q$ values perform best at large harvester sizes, $L$, and the optimal frequency $f_{opt}(L)$ also settles at $f_{peak}$. Conditional powers of harvesters considered for both locations 1 and 2, are depicted in figure 8.

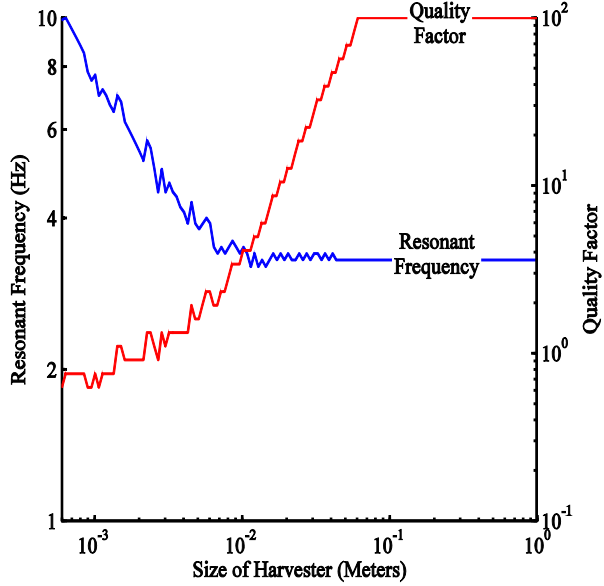
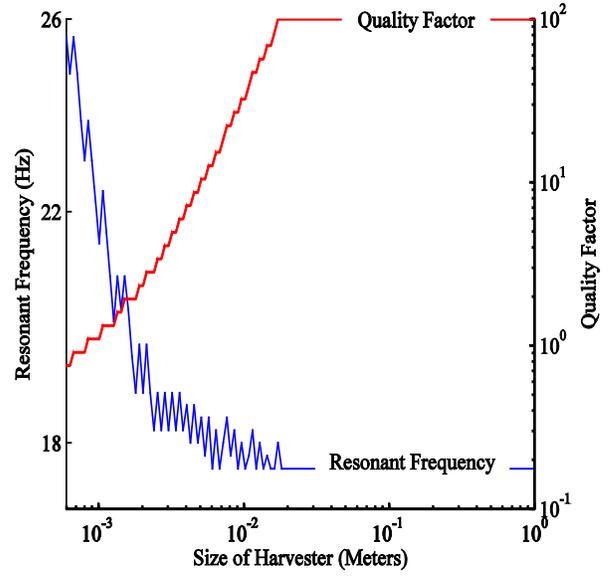

**Figure 6.** Optimum frequency and quality factor for car location 1

**Figure 7.** Optimum frequency and quality factor for car location 2.

As observed in figure 8, for location 1, the maximum power saturates to value of $P_{max} \approx 0.009 \, mW/g$, which is equal to power limit given by $S_{a,1}(f_{peak,1})/4 = 0.009 \, mW/g$, where $S_{a,1}(f_{peak,1}) \cong 0.036 \, m^2/s^3$ is evaluated at $f_{peak,1} = 3Hz$. For location 2, saturation value corresponds to power limit of $P_{max} \approx 0.12 \, mW/g$ which again exhibits excellent coherence with derived power limit, given by $S_{a,2}(f_{peak,2})/4 = 0.12 \, mW/g$, where $S_{a,2}(f_{peak,2}) \cong 0.48 \, m^2/s^3$ is evaluated at $f_{peak,2} = 20Hz$. The output power saturates at the critical harvester sizes, $L_{c,1} = (1/f_{peak,1})\sqrt{S_{a,1}(f_{peak,1})/w_1} = (1/3)\sqrt{0.036/5.2} \approx 28 \, mm$ for location 1 and $L_{c,2} = (1/f_{peak,2})\sqrt{S_{a,2}(f_{peak,2})/w_2} = (1/18)\sqrt{0.48/15.8} \approx 9.7 \, mm$ for location 2, as predicted by equation (15). Specifically, power output at $L_{c,1} = 28mm$ is about 0.0086 mW/g, which is only 4.4% lower than saturation power of $P_{max} = 0.009 \, mW/g$ for location 1. For location 2, power output of 0.114 mW/g at $L_{c,2} = 9.7mm$ is only about 5% lower than $P_{max} = 0.12 \, mW/g$. Again, we see that $L_c$ calculated using equation (15), gives a good estimate of harvester size required to obtain maximum power.

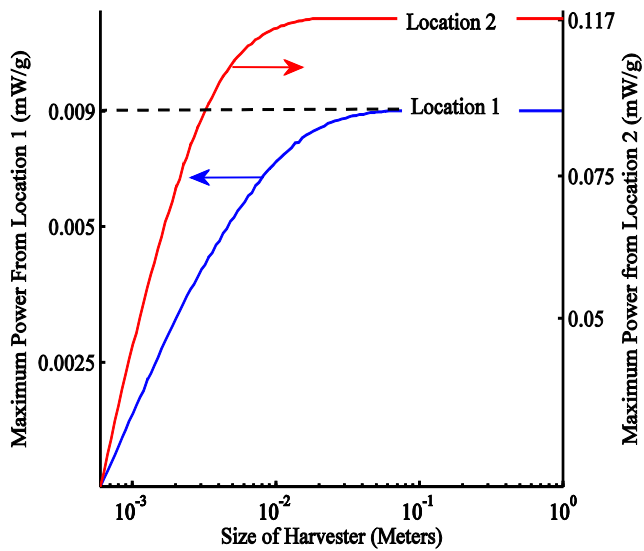
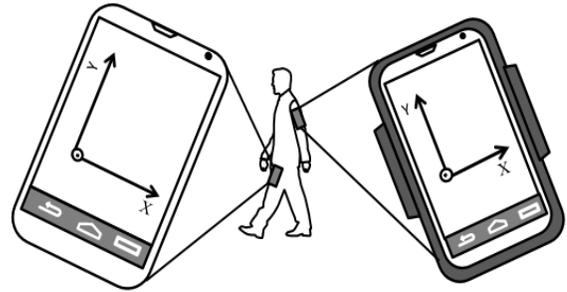

**Figure 8.** Conditional power calculated for locations 1 and 2 in the car

**Figure 9.** Nexus 4 configurations for pocket and arm based vibration measurements. The Z axis is pointing out of the paper plane towards the reader

*3.3. Walking motion*

Portable devices experience vibrations when they are carried by a walking person. Both translational and rotational vibration accelerations are present during walking and are, in fact, often generated by the same movement. For example, as the leg rotates around the hip joint during walking, it induces both rotational and translational accelerations in the devices in the hip pocket. For the purpose of this paper, we explored three different locations typical for portable electronic devices: (a) hip pocket, (b) upper arm, and (c) backpack. The time-domain vibration spectra of both linear and rotational accelerations were recorded using three-axis MEMS accelerometer and gyroscope sensors in a Google Nexus 4 smart phone. Location of the phone in cases (a) and (b) along with the nomenclature of the right handed co-ordinate system is indicated in figure 9.

Vibrations occurring in hip trouser pockets were recorded by positioning the Nexus 4 in the trouser pocket approximately vertically, such that gravity acted along the negative Y axis. As one of the co-authors walked, the acceleration signals were recorded with a sampling frequency of 50Hz for walking sessions that lasted typically for 11 minutes using Android app 'Sensor Kinetics Pro'. Subsequently, Welch averaged PSDs were obtained for all components of linear and rotational accelerations as indicated in figure 10 and figure 11, respectively. Comparing the peak values of each component of linear acceleration represented in the insets of figure 10, we notice that the highest peak corresponds to vertical acceleration component, $a_y$, and therefore restrict all further the analysis with respect to linear acceleration to this specific component. In figure 10, we see a number of peaks, with the highest PSD value of about $60 \, m^2/s^3$ at about $1.88 \, Hz$. This result agrees well with previous spectral measurements from walking gait [33, 36] and validates the recorded spectra using a smart phone accelerometer.

Variability of PSD across phone models was tested by comparing the spectra in figure 10 and figure 11 with those obtained with a Samsung Galaxy S4 smart phone. The recorded spectra were found to agree within about 10%. Accelerometer and gyroscope sensors of Nexus 4 offer excellent performance and stability during high sampling frequency operation [40]. Hence, vibration data from walking experiments were gathered using Nexus 4 sensors.

For the Z component of rotational signal we observe from figure 11, a narrow peak at around $1\,Hz$ and also a broad peak at about $7.5 - 8\,Hz$. Comparing figure 10 and figure 11, we see that the narrow peak in the rotational signal case occurs at almost half the frequency (1Hz) as that of linear acceleration signal (~2Hz). This is consistent with fact that one complete rotational oscillation of the leg corresponds to two foot strikes on ground. Although the 'X' component of rotational signal yields marginally higher peak PSD value as observed in figure 11, we choose the 'Z' component since pocket sized devices, such as phones and key fobs, are usually oriented vertically in the hip pockets. Figure 10 and figure 11 also illustrate the difference between PSDs obtained from two different walks.

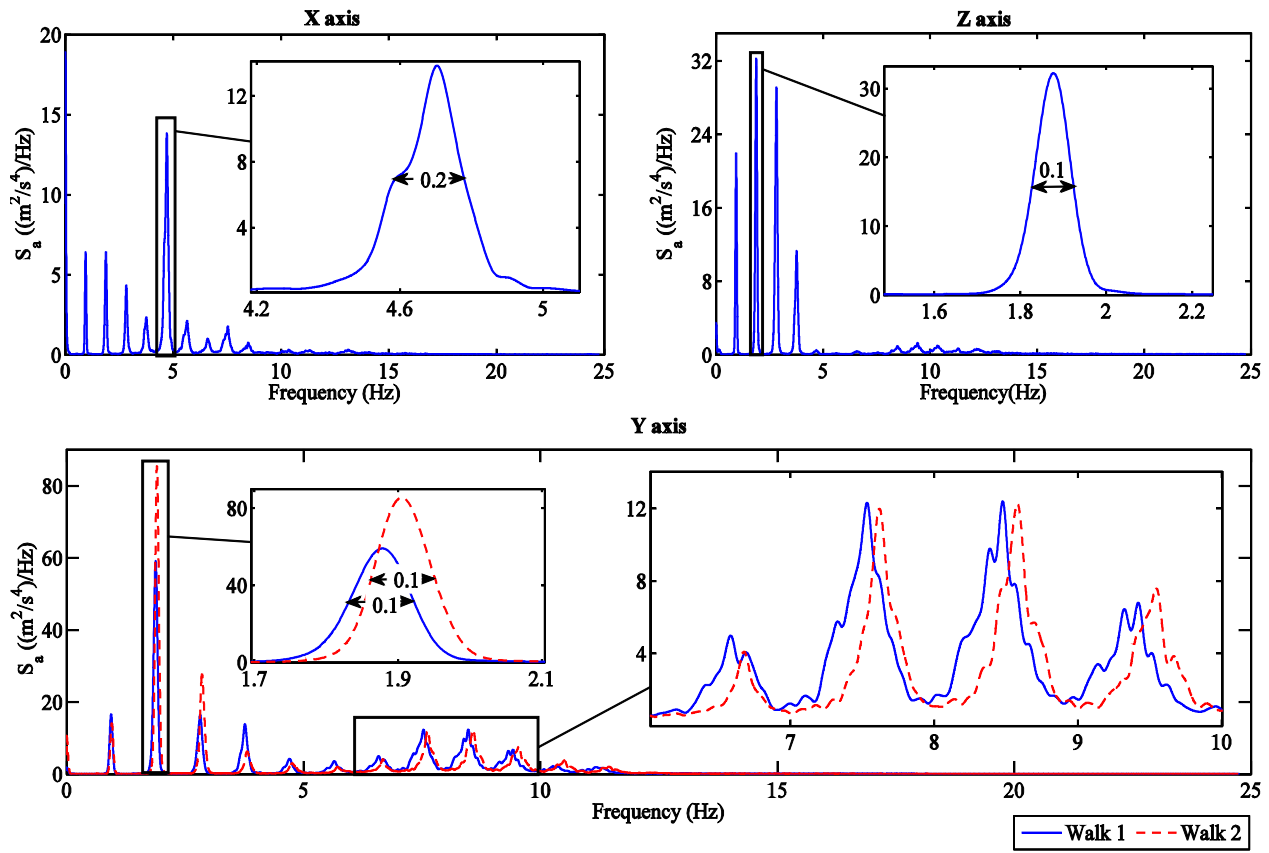

**Figure 10.** PSD of linear acceleration for a pocket harvester.

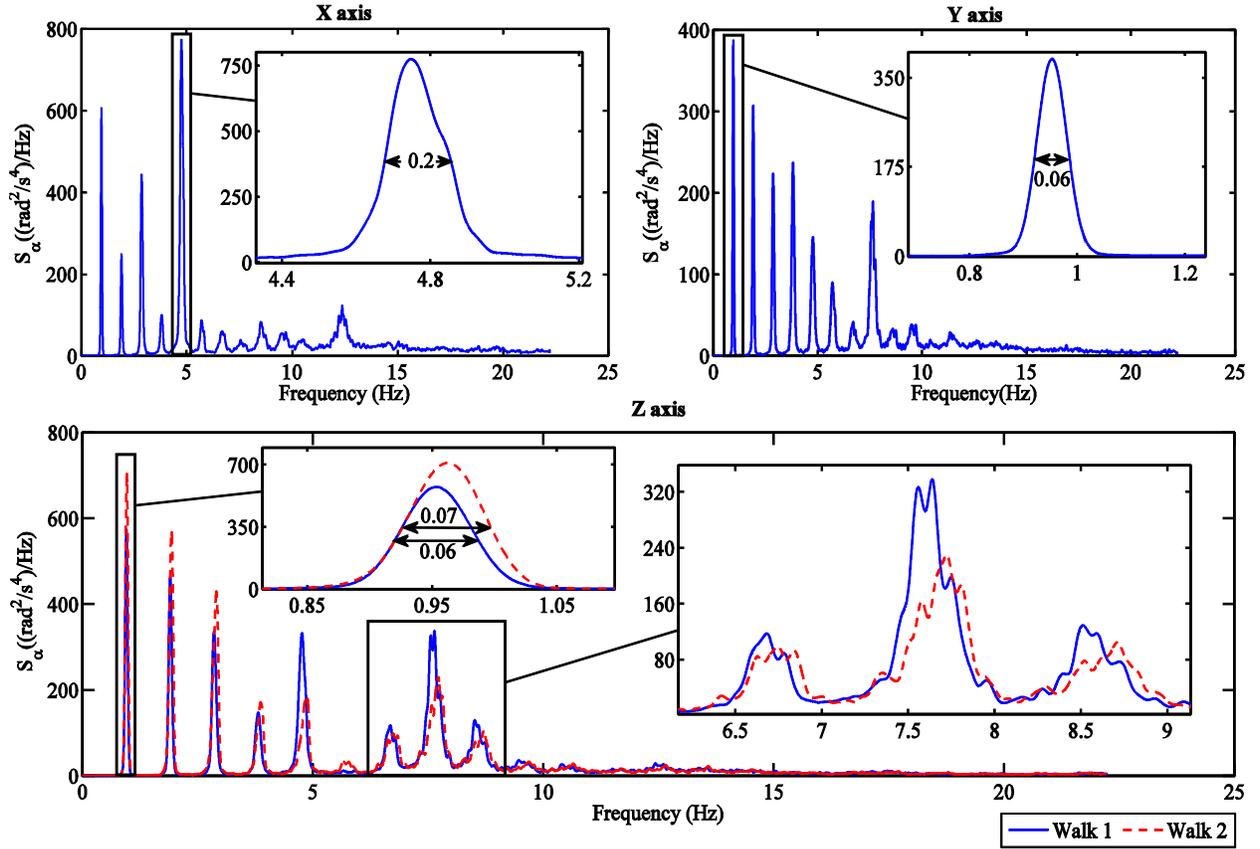

**Figure 11.** PSD of rotational acceleration for a pocket harvester.

Optimum performance parameters $f_{opt}(L)$ and $Q_{opt}(L)$ for translational harvesters which are represented in figure 12 were derived from the corresponding PSD illustrated in figure 10. Maximum power output of translational harvester was evaluated using equation (8).

For rotational harvesters, it is beneficial to adopt a ring type design which provides highest value of moment of inertia, $I$, for a given harvester size and mass. The conditional power for such a ring type harvester, $P_{c,r}(L)$, with the ring diameter $D$ and moment of inertia $I = mD^2/4$, can be calculated using equation (16) as:

$$P_{c,r} = P_r(L)/m = S_{\alpha,peak} D^2/16 \tag{18}$$

In this case, characteristic dimension of the ring, $D$, determines the harvester size, $L = D$. A comparison of power limits of these two architectures is illustrated in figure 13, which reveals that it is favorable to adopt a rotational harvester design only for sizes $L > 26\,cm$. Commonly used devices such as key fobs and phones are much smaller, with sizes of about 3 cm and 6 cm respectively, as indicated in figure 13. For such pocket-sized devices, translational harvesters are therefore the more promising architecture. Figure 13 shows that maximum power from a translational harvester saturates at about $15.1\,mW/g$, is almost exactly equal to $1/4$ times peak PSD value of $60\,m^2/s^3$ shown in figure 10. The power saturates when harvester size exceeds the critical size $L_c \approx (1/f_{peak})\sqrt{S_{a,peak}/w} = (1/1.88)\sqrt{60/0.1} \approx 13\,m$.

In particular, the conditional power at the critical size is approximately 14.3 mW/g, which is only 5.3 percent below the saturation value of 15.1 $mW/g$.

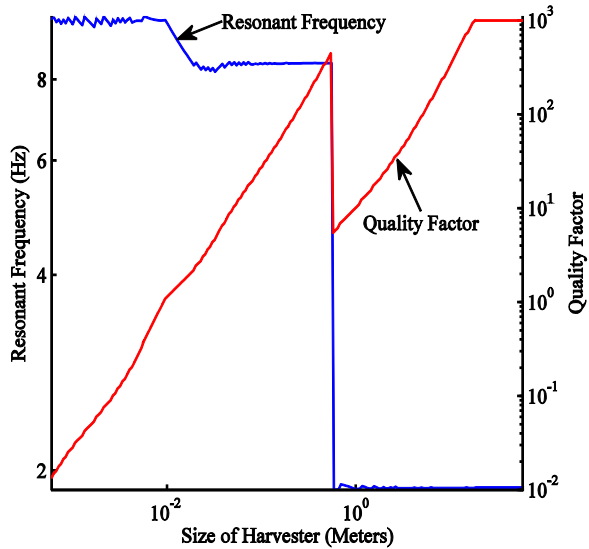
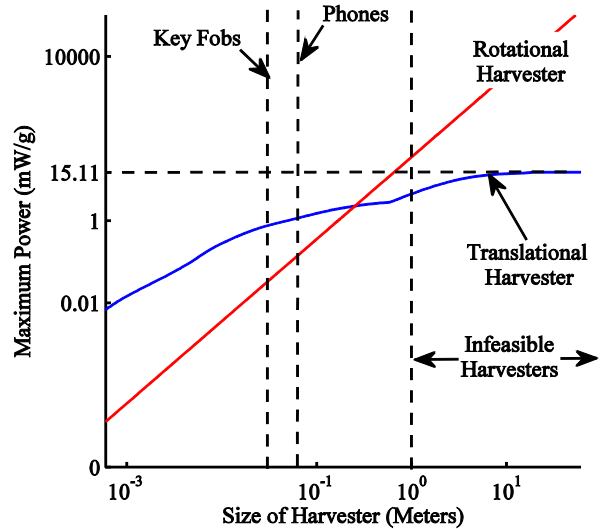

**Figure 12.** Optimum frequency and quality factor for pocket harvester.

**Figure 13.** Conditional power from pocket harvesters.

In another experiment, Nexus 4 was fitted in an armband, as shown in figure 9. Both linear and rotational acceleration signals were recorded and their respective PSD spectrums were derived as shown in figure 14. Similar to previous cases, $f_{opt}(L)$, $Q_{opt}(L)$, $P_c(L)$ and $P_{c,r}(L)$ were calculated. These are presented in figure 15 and figure 16 respectively. Although the overall trend in figure 16 is similar to that in figure 13, the crossover size beyond which rotational harvesters offer superior performance is now much lower, at $L = 8.2\ cm$. Devices such as phones and MP3 players, which are frequently mounted on upper arms during walking, jogging, *etc.*, are smaller in width than this crossover size, as indicated in figure 16. Therefore translation harvester remains the better choice resulting in peak powers of up to $20\ mW/g$. The critical size, $L_c$ for this case, as calculated from Eq.15 is about $L_c \approx (1/f_{peak})\sqrt{S_{a,peak}/w} = (1/1.88)\sqrt{74/0.07} \approx 17\ m$. At this size, the harvester power output is approximately $18\ mW/g$, which is only about 5% lower than saturation power $P_{max} = 19\ mW/g$, as seen from figure 16.

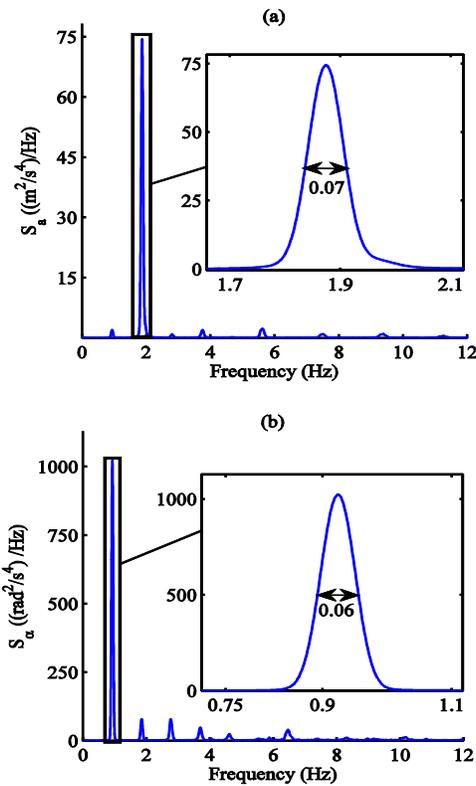

**Figure 14.** PSD of a) linear b) rotational accelerations for arm based harvesters.

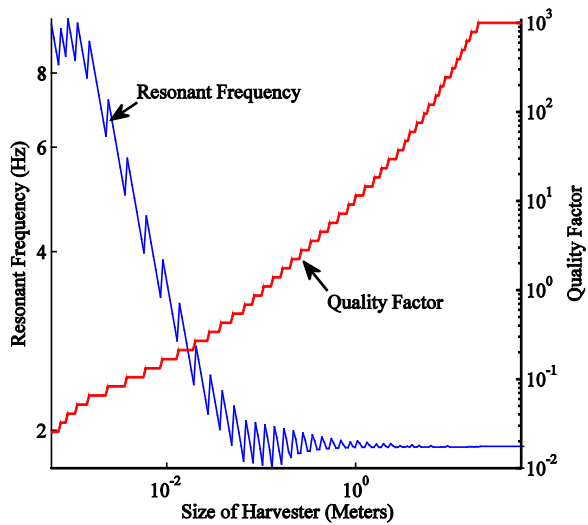

**Figure 15.** Optimum frequency and quality factor for arm based harvester.

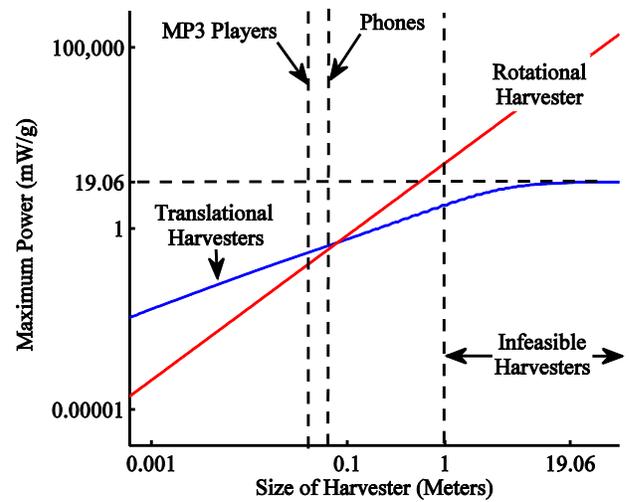

**Figure 16.** Conditional power from arm based harvesters.

In the third experiment, the harvester was fastened inside a backpack in a vertical orientation, similar to the positioning adopted for the pocket case. PSD spectrum of translational and rotational signals for this case is illustrated in figure 17. Routine analysis was then carried out, and the characteristic $f_{opt}(L)$, $Q_{opt}(L)$, $P_c(L)$ and $P_{c,r}(L)$ plots for this case are shown in figure 18 and figure 19. For this case, the crossover size was found to be about 83 cm, almost 10 times the corresponding value for upper arm case. This can be attributed to the low coupling between translational and rotational motions of human torso, as compared to its strong influence for the case of harvesters mounted on arms. Again, the crossover size is bigger than practical devices like phones, pacemakers, reiterating advantage of translational structures for realistic devices with practical size. Theoretical peak powers that can be obtained in this case were found to be the highest, about $30\ mW/g$. However, the critical harvester size for this case, evaluated using equation (15), is very large: $L_c \approx (1/f_{peak})\sqrt{S_{a,peak}/w} = (1/1.8)\sqrt{120/0.08} \approx 21\ m$. Power output at $L_c = 21\ m$ is about 29 mW/g, which is only about 3.3% lower than saturation power, $P_{max} = 30\ mW/g$. For harvesters with realistic sizes (less than 10 cm), the maximum power is of the order of 1 mW/g in all three cases we have considered (pocket, armband, and backpack). This is in agreement with the limits previously derived using time-domain simulations [36], confirming the validity of our approach.

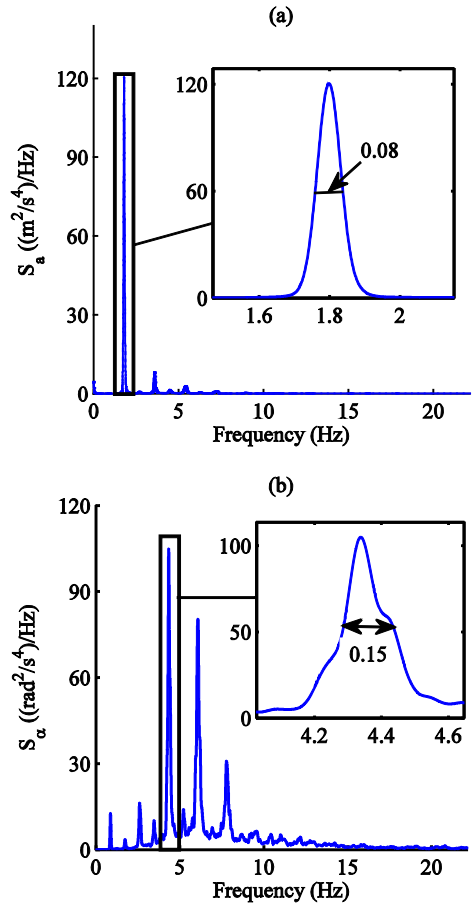

**Figure 17.** PSD of (a) linear (b) rotational accelerations for backpack-mounted harvester.

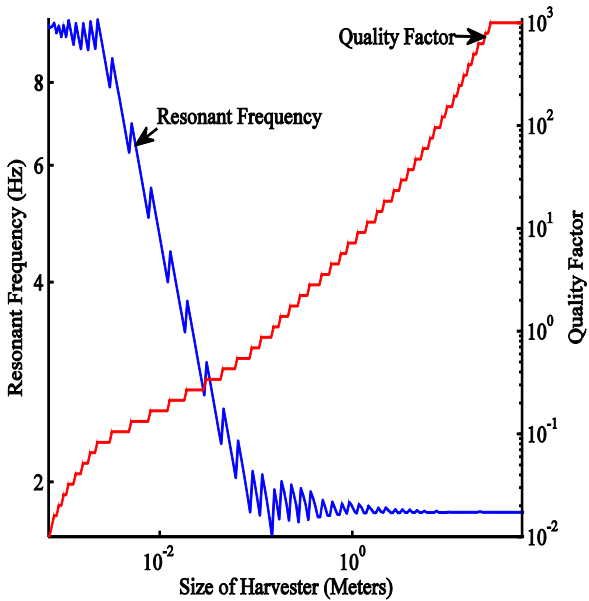 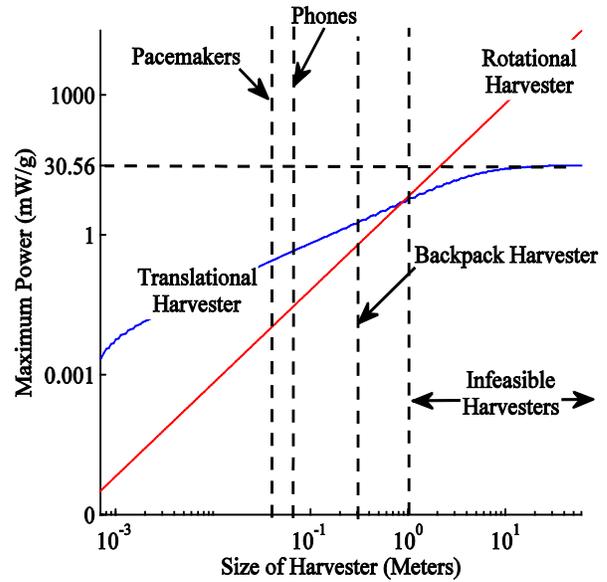

**Figure 18.** Optimum frequency and quality factor for backpack-mounted harvesters.

**Figure 19.** Conditional power from backpack-mounted harvesters.

## 4. Conclusion

This paper presents in detail the power spectral density-based technique for estimating power limits of translational and rotational inertial vibration power harvesters. We have demonstrated, for the first time, how these limits can be applied to practical situations using three example vibration sources: a small power tool, idling car, and a walking person. Also, a novel formula that estimates required harvester size for approaching these power limits was derived. We demonstrated how, using just the measured power spectral density plot, one can easily estimate the absolute highest power that a linear velocity-damped inertial harvester can achieve, as well as the minimum size of the harvester necessary to achieve this maximum power. More detailed analysis can be used to obtain the optimum resonator parameters (resonant frequency and quality factor) and maximum power as a function of the harvester size.

It was found that for each of these cases, maximum power increased with size and saturated at the expected critical sizes for large translational harvesters. In terms of the power output potential~300 mW/g could be harvested from the vibrations of a pneumatic power tool by employing a harvester with minimum size of 33 mm. In contrast, the vibrations of an idling car can yield at most 0.01 W/g for a 28 mm sized harvester located at passenger firewall (suspended) and ~0.1 mW/g for a 15 mm sized harvester located on the shock tower (unsuspended). We note that, in both the car and power tool cases, these ultimate power limits could thus be achieved with moderately sized harvesters (<10 cm). For the case of walking, the power saturated at unrealistically large sizes, on the order of 10 meters, whereas practical devices with sizes of less than 10 cm were limited to maximum output power on the order of ~ 1 mW/g. For these moderate sizes, translational harvesters outperform vibrations harvesters for all three considered locations: (a) pocket, (b) upper arm, and (c) backpack.

In summary, the approach presented in this paper can be used to obtain quick estimates of the maximum power derivable from any random (non-periodic) vibration source with as well as the corresponding harvester size, mass, and other parameters required to achieve those power levels. These estimates provide not only a quick feasibility check for vibration power harvesting in a given vibration environment, but also an insight into the optimal design and performance characteristics required to achieve this maximum power.

**Acknowledgements**
The authors would like to thank the Bosch Research and Technology Center, Andrew Pullin and Dan Steingart for their help in obtaining vibration measurement data. This work was supported by the School of Engineering and Applied Sciences at University of Pennsylvania.